\documentclass[12pt]{article}

\usepackage{amsmath,amssymb}
\usepackage{epsfig,lscape,rotating}
\usepackage{graphicx}
\usepackage{color}

\include{newcommands}

\def\beq{\begin{equation}}
\def\eeq#1{\label{#1}\end{equation}}
\def\eeqn{\end{equation}}
\def\beqa{\begin{eqnarray}}
\def\eeqa#1{\label{#1}\end{eqnarray}}
\def\eeqan{\end{eqnarray}}
\def\CR{\nonumber \\ }
\def\leqn#1{(\ref{#1})}

\def\stacksymbols #1#2#3#4{\def\theguybelow{#2}
    \def\vp{\lower#3pt}
    \def\sp{\baselineskip0pt\lineskip#4pt}
    \mathrel{\mathpalette\intermediary#1}}

\def\intermediary#1#2{\vp\vbox{\sp
     \everycr={}\tabskip0pt
     \halign{$\mathsurround0pt#1\hfil##\hfil$\crcr#2\crcr
              \theguybelow\crcr}}}

\def\gsim{\stacksymbols{>}{\sim}{2.5}{.2}}
\def\lsim{\stacksymbols{<}{\sim}{2.5}{.2}}

\def\to{\rightarrow}
\def\V{V_{\rm eff}}

\begin{document}
\begin{titlepage}

\vskip1.5cm
\begin{center}
{\huge \bf Higgs Self-Coupling as a Probe of \\ \vskip.4cm
Electroweak Phase Transition}
\vskip.8cm
\end{center}

\begin{center}
{\bf Andrew Noble and Maxim Perelstein } \\
\end{center}
\vskip 8pt

\begin{center}

\vspace*{0.2cm}

{\it Institute for High Energy Phenomenology\\
Newman Laboratory of Elementary Particle Physics, \\
Cornell University, Ithaca, NY 14853, USA } \\

\vspace*{0.3cm} {\tt
anoble@physics.cornell.edu,maxim@lepp.cornell.edu }
\end{center}

\begin{abstract}
We argue that, within a broad class of extensions of the Standard Model,
there is a tight corellation between the dynamics of the electroweak
phase transition and the cubic self-coupling of the Higgs boson: Models
which exhibit a strong first-order EWPT predict a large 
deviation of the Higgs self-coupling from the Standard Model prediction,
as long as no accidental cancellations occur. Order-one deviations are
typical. This shift would be observable at the Large Hadron Collider if
the proposed luminosity or energy upgrades are realized, as well as at
a future electron-positron collider such as the proposed International 
Linear Collider. These measurements would provide
a laboratory test of the dynamics of the electroweak phase transition. 
\end{abstract}

\vglue 0.3truecm

\end{titlepage}

\section{Introduction}

About $10^{-10}$ sec after the Big Bang, the universe underwent a phase 
transition in which the electroweak gauge symmetry was reduced from 
$SU(2)_L\times U(1)_Y$ to $U(1)_{\rm em}$. In the Standard Model (SM) and 
many of its extensions, the electroweak phase transition (EWPT) is 
understood 
in terms of the dynamics of a fundamental scalar field, the Higgs $h$: the 
vacuum $h=0$ preferred at high temperatures becomes unstable at a certain 
``critical temperature'' $T_c$ and the Higgs develops a vacuum expectation 
value (vev)~\cite{EWPT}. The nature of this transition is a subject 
of considerable interest. For example, is the EWPT a violent, out-of-equilibrium 
event accompanied by massive entropy production (first-order transition)
or a gentle, quasi-adiabatic process (second-order transition)? Answering 
this
question would have far-ranging implications: for example, electroweak 
baryogenesis, which theoretically is arguably the most attractive of the 
mechanisms proposed to explain the observed matter-antimatter asymmetry, can 
only occur if the EWPT is strongly first-order~\cite{EWbaryo}. It is very
difficult to obtain direct information on the EWPT from astrophysical 
observations, due to the long and complex history of the Universe since the 
transition. (Note, however, that gravitational waves produced in a 
first-order 
transition might be observable at LISA~\cite{Grojean:2006bp}.) 
Luckily, the dynamics of the phase transition can be deduced theoretically, 
once the
fundamental Lagrangian governing the physics at the electroweak scale is
known. For example, it is well known that if there is no physics beyond the 
SM relevant at that scale, the transition is never first-order for the
Higgs boson mass obeying the current experimental lower bound, 114.4 GeV.
Of course, the gauge hierarchy problem and other arguments strongly 
suggest that physics beyond the Standard Model (BSM) should appear at the 
electroweak scale, 
and the nature of the EWPT depends on the (as yet unknown) details of that
physics. Observing all of the weak-scale BSM particles and measuring their 
masses and interactions (in particular with the Higgs) would provide 
sufficient information to determine the order of the transition. In practice, 
however, this will be a very difficult task for the upcoming Large Hadron 
Collider (LHC), and even for a more distant next-generation $e^+e^-$ collider 
such as the proposed International Linear Collider (ILC), especially in 
models where some of the new states are gauge singlets. It would therefore 
be useful to identify simpler observables that are correlated with the 
order of the transition. In this paper, we propose that the value of the 
Higgs boson cubic self-coupling is one such 
observable. Specifically, we argue that models leading to a first-order phase transition predict a large (typically order-one) deviation of the Higgs cubic coupling from its SM value. We show that this is the case in three toy models, which capture three basic mechanisms for obtaining a first-order EWPT: loop-induced corrections to the Higgs effective potential, nonrenormalizable operators in the potential, and mixing between the weak-doublet Higgs and singlet scalars at the weak scale. The only exceptions to this statement that were observed in our analysis are due to accidental cancellations between large corrections to the Higgs cubic of different origins. Thus, while non-observation of a large deviation of the Higgs cubic from the SM prediction would not completely rule out a first-order EWPT, it would make it very unlikely.  

The correlation between a strong first-order EWPT and a large
deviation of the Higgs self-coupling from the SM value was already noted,
in the specific contexts of low-cutoff models~\cite{GSW} and the two-Higgs 
doublet model~\cite{2HDM}.
Here, we will argue that this corellation is in fact much more general, and
applies (with a small caveat requiring no accidental cancellations) in a
very broad class of models.

\section{Renormalizable Models with Unmixed Higgs}
\label{sec:2}

In this paper, we will not restrict ourselves to a specific extension of 
the SM. We will instead study a series of toy models, which capture
the important features of realistic BSM theories exhibiting strong
first-order EWPT. The toy models analyzed in this section satisfy two basic
conditions. First, the only scalar field which changes its value during 
the EWPT is the neutral component $h$ of the SM Higgs doublet\footnote{The
``SM Higgs doublet'' is defined as a field whose tree-level couplings to  
all SM states are given
by the SM values, with no large corrections from new physics. For example,
no significant tree-level mixing of $h$ with gauge-singlet scalars is 
allowed.} $H=(G^+, (h+iG^0)/\sqrt{2})$. With this assumption, the 
finite-temperature effective 
potential $V_{\rm eff}(h,T)$ completely specifies the dynamics of the EWPT. 
Second, at tree level, the Higgs sector is weakly coupled and can be 
adequately described by a renormalizable Lagrangian, i.e.~no irrelevant 
terms are important. Many theoretically motivated models satisfy these
conditions, including the Minimal Supersymmetric Standard Model 
(MSSM)~\cite{MSSM} for a broad range of parameters corresponding to the 
so-called Higgs decoupling region, Little Higgs models~\cite{LH} such as 
the Littlest Higgs with T-parity, and others. Interesting models which 
violate either one of these conditions and exhibit first-order EWPT also
exist; we will study such examples in sections~\ref{sec:3} and~\ref{sec:4}.

With the above conditions, the tree-level Higgs potential has the form 
\beq
V^t(h) = -\frac{\mu^2}{2} h^2 + \frac{\lambda}{4} h^4\,,
\eeq{Vtree}
while the masses of the SM and BSM particles are given by
\beq
M_i^2(h) = M_{0i}^2 + a_i h^2.
\eeq{masses}
At the one-loop level, the Higgs effective potential, including 
finite-temperature corrections, can be determined completely once
the values of $M_{0i}^2$ and $a_i$ for every particle in the theory are 
known. In practice, particles with $a_i\ll 1$ (weak coupling to the Higgs)
can be neglected. In the SM, these parameters are given by
\beqa
i &=& (t, W, Z, h, G)\,,\CR
M_0^2&=&(0,0,0,-\mu^2,-\mu^2)\,,\CR
a &=& (\frac{\lambda_t^2}{2}, \frac{g^2}{4}, \frac{g^2 + g^{\prime 2}}{4}, 
3\lambda, \lambda)\,.
\eeqa{SM}
The zero-temperature (Coleman-Weinberg) 
correction to the potential~\leqn{Vtree}, in the $\overline{MS}$ 
regularization scheme, has the form
\beq
V^l_0(h) = \sum_i \frac{g_i (-1)^{F_i}}{64\pi^2}\,M_i^4 (\log 
\frac{M_i^2}{\Lambda^2} + C_i)\,
\eeq{Vzero}
where $\Lambda$ is the renormalization scale and $g_i$ and $F_i$ are the 
multiplicity and fermion number of the field $i$. In the SM, in the 
basis of Eq.~\leqn{SM}, $g =(12, 6, 3, 1, 3)$ and $F=(-1,1,1,1,1)$.
The constants $C_i$ are regularization scheme-dependent and are physically
irrelevant, since their only effect is an additive renormalization of the
parameters $\mu^2$ and $\lambda$ and the cosmological constant. Rescaling
$\Lambda$ has the same effect; we fix $\Lambda=174$ GeV and interpret 
$\mu^2$ and $\lambda$ as the values of the couplings at that scale.
The parameters are subject to the constraint  
\beq
\frac{d}{dh}\left(V^t(h=v)+V^l_0(h=v)\right) \,=\,0
\eeq{vdef}
where $v=246$ GeV. In addition, the Higgs mass, 
\beq
m_h^2 = \frac{d^2}{dh^2}\left(V^t(h=v)+V^l_0(h=v)\right)\,,
\eeq{mhphys}
must satisfy the experimental lower bound, 
\beq
m_h>114.4~{\rm GeV}. 
\eeq{hbound}
The potential defined in this way suffers from an 
infrared singularity that
appears at the minimum of the tree-level potential, $v_{\rm tree}^2 = 
\mu^2/\lambda$, and is due to the massless (Goldstone) 
bosons $G$ at that point~\cite{AH}. While the potential itself remains finite
($\lim_{x \to 0} x^4 \log x^2 = 0$), the cubic coupling defined 
in~\leqn{cubic} diverges at that point. Resummation of the infrared 
logarithms is required to correctly evaluate $\lambda_3$ in the neighborhood 
of that point. This resummation is conveniently achieved by switching
from the zero-momentum renormalization scheme to the on-shell 
scheme~\cite{AH}. Technically, this is equivalent to adding an extra term to
$V^l_0$, and interpreting the second derivative at the minimum of the 
corrected potential as the physical (on-shell) Higgs mass. The additional
term eliminates the infrared divergence. We will employ the on-shell scheme
throughout this paper.

The one-loop finite-temperature potential has the form~\cite{WDJ} 
\beqa
V_T(h, T) &=& \sum_{F_i=0} \frac{g_iT}{2\pi^2}\int dk k^2 
\log[1-\exp(-\beta\sqrt{k^2+M_i^2(h)})] \CR & &- \sum_{F_i=1} 
\frac{g_i T}{2\pi^2} \int dk k^2 \log[1+\exp(-\beta\sqrt{k^2+M_i^2(h)})],
\eeqa{VT}
where $\beta = 1/T$. If the model contains bosons with masses 
small compared to $T$, multi-loop infrared-divergent contributions need to be 
included to correctly analyze the dynamics of $h$. These contributions can be 
resummed and written in the form of the ``ring 
potential''~\cite{Carrington:1991hz}: 
\beq
V_r(h, T) = \sum_{i} \frac{T}{12\pi}\, {\rm Tr}\left[M_i^3(h) - 
(M_i^2(h)+\Pi_i(0))^{3/2}\right],
\eeq{Vr}
where the sum runs over all (light) bosonic degrees of freedom, and 
$\Pi_i(0)$ is the zero-momentum polarization tensor. In the SM,  
\beqa
\Pi_h(0) \,=\, \Pi_G(0) &=& T^2 \, \left(\frac{3}{16}g^2+\frac{1}{16}
g^{\prime 2}+\frac{1}{4}\lambda_t^2+\frac{1}{2}\lambda \right)\,,\CR
\Pi_{GB}(0) &=&
\frac{11}{6}T^2 \,{\rm diag}\,\left(g^2,g^2,g^2,g^{\prime 2}\right)\,,
\eeqa{SMrings}
and
\beqa
M_{GB}^2(h)\,=\,
\frac{h^2}{4}\left(\begin{array}{cccc}
g^2& 0& 0& 0\\
0& g^2& 0& 0\\
0& 0& g^2& -g g^{\prime 2}\\
0& 0& -g g^{\prime 2}& g^{\prime 2}\\
\end{array}\right).
\eeqa{MGB}
These expressions are valid at $T\gg M_W$; for smaller $T$, the ring terms 
are unimportant and can be dropped. The full 
potential\footnote{For some values of $h$, the potential $V(h,T)$ develops
an imaginary part, due to the quantum instability of a classical state with 
spatially uniform value of $h$~\cite{WW}. 
This effect does not affect the dynamics of a first-order transition, as
long as the rate of this instability is small compared to the 
electroweak scale.
This condition is always satisfied in our analysis.} 
is $\V(h; T) = V^t + V^l_0 + V_T+V_r$. A useful 
``decoupling'' feature of this potential is that at any $T$, the 
contributions 
from particles with $M_i(h)\gg T$ are exponentially suppressed. Thus, in BSM 
theories, only weak-scale states need to be included to analyze the EWPT.

At high temperature, $T\gg M_{0i}$ for all states, the potential has the 
simple form
\beq
\V(h; T) \approx \left( \sum_{F_i=0} g_i a_i  + 
\frac{1}{2} \sum_{F_i=1} g_i a_i \right)\, \frac{T^2h^2}{24}.
\eeq{highT}
If the quantity in brackets is positive, $\left<h\right>=0$ is a stable 
minimum, and the electroweak symmetry is restored. This is the case, for
example, in the SM and the MSSM, where $a_i>0$ for all states. We will 
restrict our analysis to such models, since otherwise the EWPT does not occur.
A first-order transition occurs if $\V(h; T)$ develops a second, 
symmetry-breaking local minimum, $\left<h\right>\equiv v_T(T)\not=0$, 
while the 
high-temperature minimum $\left<h\right>=0$ is still classically stable. 
Then, once the symmetry-breaking vacuum becomes energetically preferred, 
bubbles of the symmetry-breaking vacuum will start rapidly nucleating, 
growing, and coalescing, eventually encompassing the entire Universe and
completing the phase transition. The critical temperature $T_c$ at which 
the phase transition occurs can be found from the condition\footnote{Note 
that the
actual transition temperature may be lower if the bubble nucleation rate 
$\Gamma$ is strongly suppressed compared to its natural value, 
$\Gamma\sim v$. 
Phase transition does not occur until $\Gamma\sim H \sim v^2/M_{\rm Pl}$. 
Given the large hierarchy $v\ll M_{\rm Pl}$, we will not include this 
possibility in our analysis.}
\beq
\V(\left<h\right>=0; T_c)\,=\,\V(v_T(T_c); T_c). 
\eeq{Tc}
The ``strength'' of the first-order transition (more precisely, the degree of 
deviation from quasiadiabatic evolution) can be characterized by the 
dimensionless parameter
\beq
\xi = \frac{v_T(T_c)}{T_c}.
\eeq{xidef}
For example, it is well known that successful electroweak baryogenesis 
requires $\xi\gsim 1$.  

Due to the complicated form of the potential, the problem of finding 
$T_c$ and 
$v_T(T_c)$ in a given model is in general not tractable 
analytically. We have developed a numerical code which is used to evaluate
these quantities in all examples studied in this paper. 

In the SM, the EWPT is always second-order once the bound~\leqn{hbound} is
satisfied. Below, we analyze the situation in a few simple extensions of
the SM.

\subsection{Single BSM Scalar} 
\label{singleS}

Consider a toy model in which a single real scalar 
state is added to the SM, with the potential
\beq
V = \frac{1}{2} M_0^2 S^2 + \zeta^2 |H|^2 S^2.
\eeq{spot}
This is the most general renormalizable potential which respects a 
symmetry $S\to -S$. (Without this symmetry, tree-level mixing between $h$
and $S$ would generally occur. Such models will be discussed below.)
Note that while we use the SUSY-inspired notation $\zeta^2$ for the 
coefficient in Eq.~\leqn{spot}, this coefficient can in general be negative,
as long as $S$ does not become tachyonic until $h> v$ and the lifetime
of the vacuum at $(h=v, S=0)$ is larger than the present age of the 
universe. 
(The runaway direction can be stabilized by the addition of a small $S^4$ 
term, which does not affect our analysis.) 

\begin{figure}[t]
\begin{center}
\includegraphics[width=7cm,keepaspectratio=true]{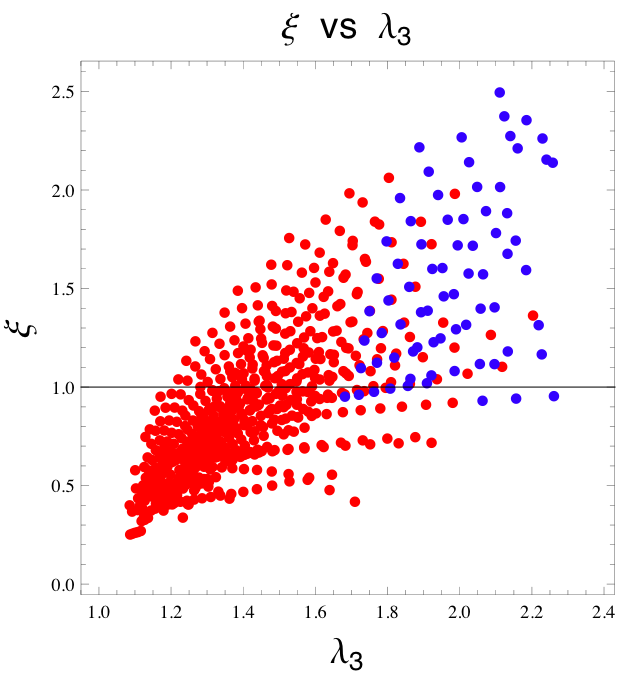}
\hspace{1cm}
\includegraphics[width=7cm,keepaspectratio=true]{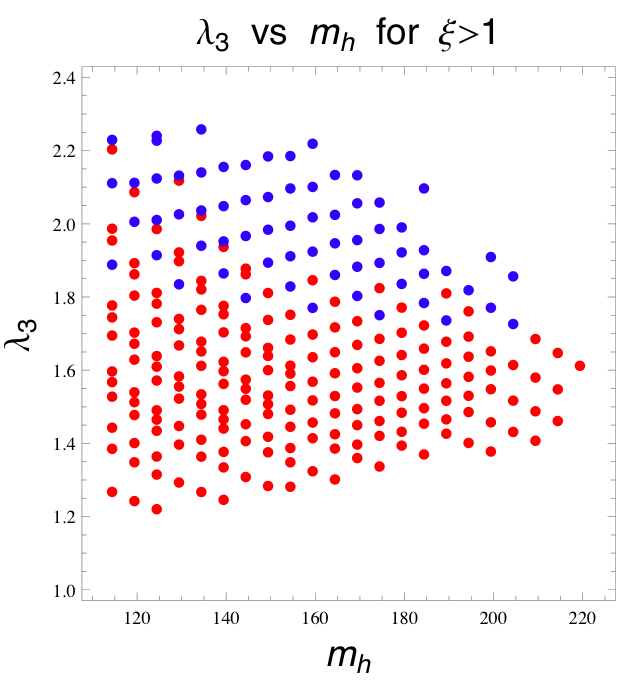}
\caption{SM with a single extra scalar. Models with a ``bumpy'' 
zero-temperature Higgs potential are shown in blue, and those without the
bump in red. (Left panel) The strength of the first-order EWPT $\xi$, defined in 
Eq.~\leqn{xidef}, vs. Higgs cubic self-coupling. (Right panel)
Higgs cubic self-coupling vs. Higgs mass for points exhibiting a strong
first-order EWPT, $\xi>1$. In both plots, the Higgs self-coupling is 
normalized to the one-loop SM expectation for the same $m_h$.}
\label{fig:scalar}
\end{center}
\end{figure}

\begin{figure}[t]
\begin{center}
\includegraphics[width=7cm,keepaspectratio=true]{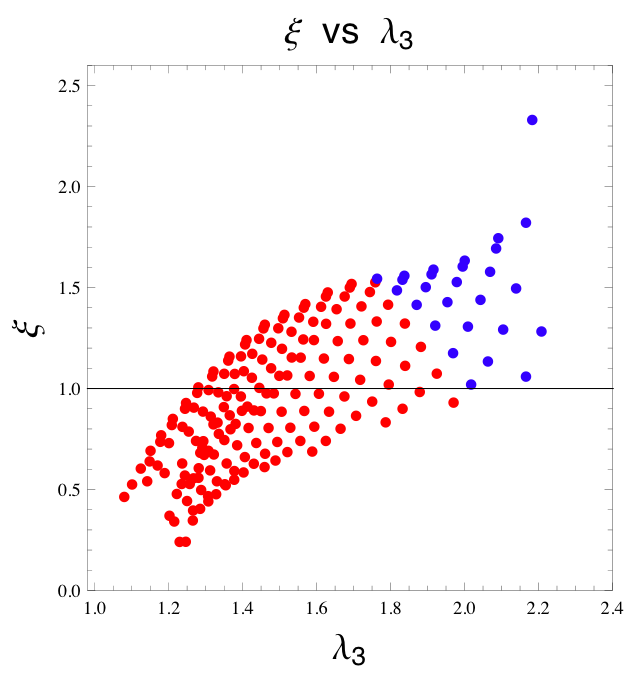}
\caption{Same as left panel of Fig.~\ref{fig:scalar}, but with the Higgs mass fixed at 160 GeV.}
\label{fig:fixedmH}
\end{center}
\end{figure}

The scalar contribution to the one-loop effective potential is given by 
Eqs.~\leqn{Vzero},~\leqn{VT},~\leqn{Vr}, where $M(S)=M_0^2+\zeta^2 h^2$,
$\Pi_S(0) = \frac{1}{3}T^2\zeta^2$ and $g_S=1$. In addition, the ring terms
due to the $h$ and $G$ fields have to be modified by adding the scalar 
contribution to the polarization tensors, $\Delta \Pi_h(0) =\Delta \Pi_G(0) 
= \frac{1}{12}T^2\zeta^2$. (We neglect additional contributions to 
polarization tensors due to $S$ self-coupling or coupling to other 
BSM singlets.) In order to identify the points with a first-order EWPT, 
we scanned the parameters of the model:\footnote{We include the models with 
negative values of $\zeta^2$ in the
scan in order to demonstrate that they exhibit the same correlation between
the Higgs cubic coupling and the $\xi$ parameter as the positive-$\zeta^2$
models. We imposed the constraint that $S$ should not be tachyonic in the 
range $h=0\ldots 300$ GeV, but did not analyze the lifetime of the 
conventional EWSB vacuum for these models, so some of the negative-$\zeta^2$ 
points may be phenomenogically unacceptable. A detailed analysis of this
issue is beyond the scope of this paper, and its outcome would not affect
the main point of our argument in any way.}  $M_0=0\ldots 800$ GeV, 
$\zeta^2=-4\ldots 8$. We also 
scanned the parameters $\mu$ and $\lambda$, making 
sure that the constraints~\leqn{vdef} and~\leqn{hbound} are satisfied; the
scanned region corresponds to $m_h=114.4\ldots 244.4$ GeV. (We did not find 
any points with higher $m_h$ and a strong first-order EWPT.) For
every point exhibiting a first-order EWPT, we compute the one-loop value of
the physical Higgs boson cubic self-coupling,
\beq
\lambda_3 \equiv \frac{1}{6}\,\frac{d^3(V^t(h)+V^l_0(h))}{dh^3}|_{h=v}.
\eeq{cubic}
The results are presented in Fig.~\ref{fig:scalar}. In the left panel, we 
plot the parameters $\xi$ and $\lambda_3$ (normalized to the SM one-loop value)
for the points in our scan; in the right panel, we plot $m_h$ and 
$\lambda_3$. It is clear from the plots that the Higgs self-coupling  
is significantly larger than the SM value for {\it all} points with a
first-order EWPT in this model. If $m_h$ is known from experiment (as it surely will be, by the time the experiments become sensitive to the Higgs self-coupling), the correlation between $\lambda_3$ and $\xi$ becomes tighter, as illustrated in Fig.~\ref{fig:fixedmH}. 

Experimental measurement of $\lambda_3$ requires collecting a significant 
sample of events with two Higgs bosons in the final state. At the LHC, 
this coupling can be determined with 20-30 \% precision for the Higgs mass
between 160 and 180 GeV, for integrated luminosity of 3000 fb$^{-1}$, which
can be achieved if the luminosity upgrade to $10^{35}$ cm$^{-2}$s$^{-1}$ is
realized~\cite{BPR1}. If an energy upgrade to a 200 TeV VHLC is realized, the 
coupling can be measured with $8-25$\% precision for 150 GeV $< m_h < 
200$ GeV~\cite{BPR2}. For lower Higgs masses, the VLHC could make a similarly 
precise measurement of $\lambda_3$ using rare decays~\cite{BPR3}. Also,   
the 500-GeV ILC is expected to measure $\lambda_3$ 
with $20\%$ accuracy for $m_h<140$ GeV and a luminosity of $1$ 
ab$^{-1}$~\cite{Djouadi:2007ik}. Our analysis shows that if
a strong first-order EWPT occurs due to an extra scalar contribution at one  
loop, and the Higgs is sufficiently light,
these experiments will discover a deviation of $\lambda_3$ from the SM 
prediction. Higher center-of-mass energy would be required if $m_h$ is larger;
note, however, that strong first-order EWPT in this setup requires 
$m_h\lsim 250$ GeV, so a modest increase in $\sqrt{s}$ would be 
sufficient to cover all of the interesting parameter space.

In order to better understand this result, consider our model with
$\mu=M_0=0$. In this case, the one-loop Higgs potential at zero 
temperature has the form
\beq
V_0 \approx \frac{1}{4}\lambda h^4 + \frac{A}{64\pi^2}h^4 \log 
\frac{h^2}{\Lambda^2}\,,
\eeq{VhDT}
where $A$ can be expressed in terms of the coupling constants of the model, 
and the contributions from Higgs loops have been neglected. If $A>0$
(which occurs for sufficiently large $\zeta^2$ in our model), this 
potential has a stable symmetry-breaking minimum at 
$h= \Lambda \exp(-8\pi^2\lambda/A-1/4)$, and a shallow local maximum at 
$h=0$. This is of course just a realization of the purely radiative EWSB 
via dimensional transmutation (DT) {\it a la} Coleman and Weinberg~\cite{DT}. 
If now a small 
{\it positive} Higgs mass term $\mu^2>0$ is added, the potential around 
$\phi=0$ is dominated by this mass term, and the maximum at the origin is 
turned into a minimum. At the same time, if $\mu^2$ is sufficiently small, 
the symmetry-breaking minimum is preserved, resulting in a 
potential with a ``bump'' between the two minima. A bumpy zero-temperature
potential essentially guarantees that the EWPT will be of the first order,  
and most of the points with strong first-order EWPT in our scan (the blue
points in Fig.~\ref{fig:scalar}) have this feature. The Higgs self-coupling 
for the pure DT potential, Eq.~\leqn{VhDT}, is 66\% larger than the 
(tree-level) SM value\footnote{The same relation holds in any model with 
approximate conformal symmetry in the Higgs sector broken by nearly marginal 
operators~\cite{GGS}.} for the same $v$ and $m_h$:
\beq
\lambda_3^{\rm DT} = \frac{5m_h^2}{6v} \,=\, \frac{5}{3}
\lambda_3^{\rm SM,tree}\,,
\eeq{l3DT} 
which is roughly at the center of the range for the models with strongly
first-order EWPT in our
scan. Even without the bump (red points in Fig.~\ref{fig:scalar}),
the models with $\xi\geq 1$ have zero-$T$ potentials whose characteristic
feature is a flatter maximum at $h=0$ compared to the SM. This shape is
close to the DT potential, again resulting in large deviations of 
$\lambda_3$ from the SM prediction.

The connection with the pure DT case points to a potential concern about our 
model: In the DT case, the physical Higgs mass is given by 
$m_h = A^{1/2}v/(4\pi)$, so that large values of $A$
are required to satisfy the experimental lower bound on $m_h$.
Correspondingly, in our scans, fairly large values of $\zeta^2$ (between 3 
and 10) are  required to 
obtain phenomenologically consistent points with a strong first-order EWPT.
For $\zeta^2\sim 10$, one might worry about the validity of perturbation
theory that was used in our analysis. Even for more moderate $\zeta^2= 3 
\ldots 5$, two-loop corrections might be numerically significant. This is 
the case, for example, in the MSSM, where the stops effectively contribute 
with $\zeta_{\rm eff}^2=\sqrt{12}\lambda_t^2 \approx 3.5$ (see 
section~\ref{multi}). 
On the other hand, the main outcome of our analysis is not the precise 
values of $\lambda_3$ for each point in the scan, but the fact that the 
deviation from the SM in this quantity is large for models with strong
first-order EWPT. It seems very unlikely that the higher-loop 
corrections would conspire to cancel the large one-loop correction to 
$\lambda_3$ and bring it back precisely to the SM value, so our main 
qualitative conclusion should continue to hold even at strong coupling.

It should be noted that it is possible to have a large deviation of $\lambda_3$ from the SM prediction {\it without} a strongly first-order transition. We found such points in our scan. These points have values of $m_0$ above, but not too far from, the Higgs vev $v$, and large values of $a$. The zero-temperature effective potential receives a correction of order $a(v/m_0)^2$, giving a substantial correction to $\lambda_3$; the correction to the finite-temperature potential decouples exponentially, as $\exp(-m_0/T)$, and is not sufficient to yield a first-order transition. Thus, while an observation of a large correction to $\lambda_3$ would keep the door open to the possibility of a strongly first-order phase transition scenario, it would not prove it.

\if
In our analysis, however, the minimum of the one-loop
potential $v$ is typically sufficiently far away from the tree-level 
minimum due to large one-loop renormalization by top and scalar loops,  
and the $G$ contribution to $\lambda_3$ is numerically small, of order a
few percent. We discard the points in our scan where the one-loop $G$ 
contribution to $\lambda_3$ is 10\% or larger, since the validity of 
perturbation theory is questionable for these points. No such points 
appeared in the scan of this section, but this constraint will play a
(minor) role in the analysis of section~\ref{SFpair} below.
\fi

\subsection{Single BSM Fermion} 

Consider the SM with a single additional Weyl fermion $\chi$, with a 
Majorana mass term
\beq
\left(M_0 + \frac{H^\dagger H}{\Lambda} + \ldots \right) \chi \chi\,,
\eeq{Mmass}
where dots denote higher-order nonrenormalizable terms. The mass has the
form~\leqn{masses} with $a=M_0/\Lambda$, and the formalism presented above
applies, with $g_\chi=2$. Adding a single fermion also modifies the ring 
contributions: $\Delta \Pi_h(0)=\Delta \Pi_G(0)=\frac{g}{24}T^2 a$, where 
$a$ is the same coefficient appearing in Eq.~\leqn{masses}, and $g=2(4)$ for a 
Weyl (Dirac) fermion. We neglect additional ring contributions that would 
arise if $\chi$ coupled to other BSM states. In this model, we performed the
scan with $m_h=114.4\ldots 414.4$ GeV, $M_0=0\ldots 800$ GeV, and $a=-4\ldots 8$.  For $a>0$, $\lambda_3$ is always suppressed with respect to the SM value, while for $a<0$, $\lambda_3$ is enhanced. (In the $T=0$ effective potential, an $a<0$ fermion behaves like an $a>0$ boson, and vice versa.)  But in neither case do we find a strong ($\xi>1$) first-order EWPT.


Since the only inputs from the model into the EWPT analysis are the 
coefficients 
in the mass formula~\leqn{masses}, the same analysis applies in a model with a 
single Dirac BSM fermion (with $M_0=0$ if the fermion is chiral, and 
possibly $M_0\not=0$ if it is vector-like). The only change is the new 
multiplicity factor, $g_\Psi=4$. This does not affect the conclusions. 

\subsection{A Single BSM Scalar-Fermion Pair}
\label{SFpair}

\begin{figure}[t]
\begin{center}
\includegraphics[width=7cm,keepaspectratio=true]{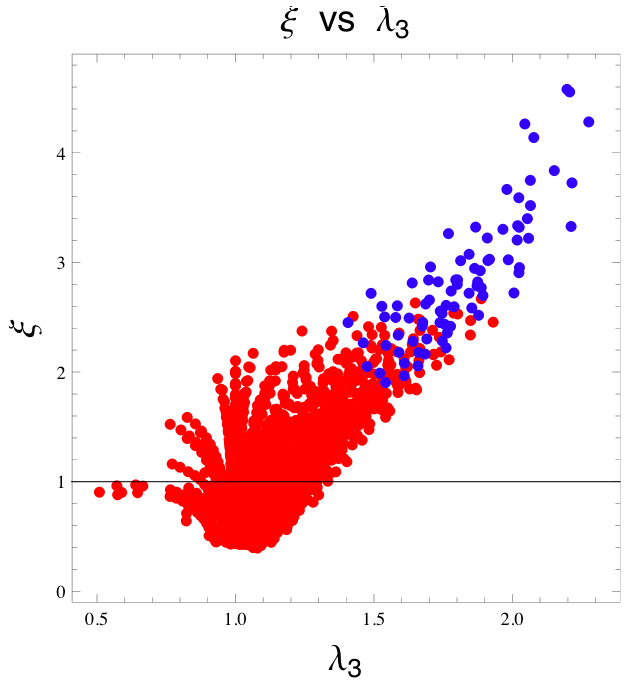}
\hspace{1cm}
\includegraphics[width=7cm,keepaspectratio=true]{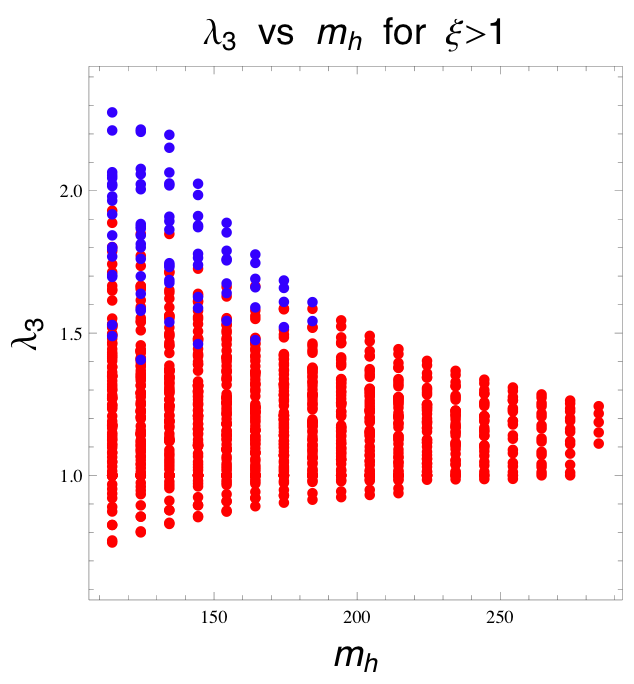}
\caption{SM with a single extra scalar-fermion pair. Models with a ``bumpy'' 
zero-temperature Higgs potential are shown in blue, and those without the
bump in red. (Left panel) The strength of the first-order EWPT $\xi$, defined in 
Eq.~\leqn{xidef}, vs. Higgs cubic self-coupling. (Right panel)
Higgs cubic self-coupling vs. Higgs mass for points exhibiting a strong
first-order EWPT, $\xi>1$. In both plots, the Higgs self-coupling is 
normalized to the one-loop SM expectation for the same $m_h$.}
\label{fig:sf}
\end{center}
\end{figure}

Any realistic extension of the SM has 
multiple bosonic and fermionic states, whose contributions to the Higgs 
effective potential must be added. In this subsection, we will explore
how the interplay between bosonic and fermionic contributions affects the
connection between the Higgs cubic coupling and the strength of a first-order
EWPT. To address this, consider a toy model with a single fermion
(for concreteness, we assume it to be a vector-like Dirac fermion),  
and four identical scalars. The number of scalars is chosen so that the 
number
of bosonic and fermionic degrees of freedom is the same, as would be the
case in supersymmetric models. We searched for points with strong first-order
EWPT in the following parameter space: $a_s=a_\Psi=-1 \ldots 4$, 
$M_0^S = 0\ldots 600$ GeV, 
$M_0^\Psi = 0\ldots 600$ GeV, $m_h=114.4\ldots 294.4$ GeV. (We did not find any points with a strong first-order EWPT for higher $m_h$.) The results are
shown in Fig.~\ref{fig:sf}. In most of the parameter space, the correlation 
between $\xi$ and $\lambda_3$, observed above in the purely scalar model,
persists, and models with strong first-order EWPT predict observable enhancements of 
$\lambda_3$ with respect to the SM value. There is also a smaller region 
where an observable suppression of $\lambda_3$ is predicted; this is due to
the dominance of the fermion loops, which contribute to $\lambda_3$ 
with a negative sign, for $M_0^\Psi<M_0^S$. There is
also, however, an exceptional parameter region, in which $\lambda_3$ is 
close to the SM value. This region has the following origin. In the one-loop 
zero-temperature potential, fermion and scalar contributions cancel 
if $M_0^S = M_0^\Psi$ and $a_s=a_\Psi$, since $S$ and $\Psi$ form an 
exact supermultiplet, and $\lambda_3$ has its SM value.
The scalar and fermion 
contributions to the finite-temperature part of the potential, however, do 
not cancel in this limit. In particular, if $a_s$ is
large, a large cubic term can be generated at high temperature, which can
lead to a first-order transition. Fermion loops do not contribute to this 
term. This explanation highlights the exceptional nature of the region with 
SM-like $\lambda_3$: the relations $M_0^S = M_0^\Psi$ and $a_s=a_\Psi$
cannot be enforced by a symmetry other than SUSY. Given that the
$S$ and $\Psi_S$ fields have significant couplings to the Higgs sector, and 
that SUSY must be broken in that sector, these relations are not radiatively 
stable. For example, renormalization group flow will naturally induce a 
mass splitting between $S$ and $\Psi$ of order
\beq
\delta M^2 \,= |M^2(S) - M^2(\Psi)| \sim \frac{\zeta^2}{16\pi^2}\,\delta 
m_h^2 \,\log \frac{M_0^2}{\Lambda^2}\,,
\eeq{deltaM}
where $\delta m_h$ is the Higgs-higgsino mass splitting (typically of order 
100 GeV or above) and $\Lambda > 1$ TeV is the messenger scale where  
logarithmic divergences are cut off. Given the large value of $\zeta^2$ in the
region where the first-order transition is possible, we expect $\delta M$
to be naturally large, of order a few hundred GeV. Imposing that the mass 
splitting be at least 100 GeV eliminates the region with first-order EWPT
and no deviation in $\lambda_3$. We find that the minimal deviation in the
cases $\delta M=(100, 200, 300, 500)$ GeV, is 7\%, 10\%, 17\% and 20\%, 
respectively. We conclude that in the absence of accidental degeneracies in
the mass spectrum, the $\lambda_3-\xi$ corellation works well in this model.

\subsection{Multiple BSM States}
\label{multi}

So far, we have observed the correlation between $\lambda_3$ and $\xi$
in simple toy models. Will it persist in realistic BSM theories with 
more complicated spectra and interactions? While it is clearly impossible 
to analyze all possible cases, it seems very likely that the answer is 
positive. Several arguments point in this direction.

First, note that in a model with $N$ real (or $N/2$ complex) scalar 
fields with identical masses and couplings, the zero-temperture
potential (and therefore $\lambda_3$) is the same as that for an 
``effective'' 
single real scalar with 
\beq
M_{0,{\rm eff}}^2 = \sqrt{N} M_0^2,\, \zeta_{\rm eff}^2 = \sqrt{N}\zeta^2
,\, g_{\rm eff}=1.
\eeq{effective}
The same scaling applies for the scalar contribution to the 
finite-temperature 
potential (including rings), except one has to also rescale $T\to N^{1/4}T$. 
If the scalar contribution dominates, this implies that the critical 
temperature
in the $N$-scalar model is reduced by $N^{1/4}$ compared to the effective
single-scalar model, and $\phi_c$ is unchanged. We checked numerically that 
this scaling is accurate to within 20\% even after the SM contributions are 
added. Thanks to this scaling, the approximate correlation between $\xi$ and 
$\lambda_3$ 
is preserved for $N>1$, and the models with $N>1$ and strong first-order
EWPT predict {\it larger} minimal enhancement in $\lambda_3$ compared to 
the $N=1$ case. The same approximate scaling laws apply to models with $N$ 
identical species of fermions as well. Since the gauge charges of the 
added states have no effect on the Higgs potential in the one-loop 
approximation, we conclude that the results of our toy-model analyses 
in fact apply to {\it any} model where a single multiplet (or a single 
supermultiplet) 
dominates the BSM contribution to the Higgs potential. The MSSM 
with light stops, and models with multiple gauge-singlet scalars as in 
Ref.~\cite{EQ}, are in this category. 

Even if multiple BSM states with different masses and couplings contribute 
to the Higgs potential, the
qualitative picture should remain the same. In fact, we checked that 
a toy model with two scalars, whose masses and Higgs couplings are varied 
independently, produces results very similar to the single-scalar case. 
This is reasonable, since the similarity between the properties of the 
models with strong first-order EWPT and the DT model, noticed in
section~\ref{singleS}, should persist independently of the precise origin 
of the large radative corrections to the Higgs potential.

Based on this evidence, we believe that significant deviation of the 
Higgs cubic self-coupling from the SM value is a generic 
prediction of weakly-coupled models of EWSB with strong first-order EWPT, provided that 
no accidental cancellations occur. A measurement of this coupling can serve 
as a diagnostic tool to glean information about the nature of the EWPT.
This would be particularly valuable in models with extra gauge singlets, 
whose masses and couplings to the Higgs are difficult to measure 
directly, especially if the decay $h\to SS$ is kinematically forbidden.

\section{A Model with Nonrenormalizable Higgs Interactions}
\label{sec:3}

The effect of new physics at the scale $\Lambda\gg v$ on the Higgs potential
can be parametrized by adding nonrenormalizable operators, suppressed by
powers of $\Lambda$, to the potential. Grojean, Servant and Wells (GSW) 
pointed out that in the presence of such terms, strong first-order EWPT
may occur~\cite{GSW,HK,DGW}. Following GSW, consider a model with a tree-level 
Higgs potential of the form
\beq
V^t(H) \,=\, \mu^2 H^\dagger H \,+\, \lambda (H^\dagger H)^2 \,+\, 
\frac{1}{\Lambda^2} (H^\dagger H)^3\,.
\eeq{GSW_Vh}
It is reasonable to use this effective theory to describe the EWPT if the
electroweak symmetry-breaking vev is small compared to $\Lambda$. 
The strong first-order EWPT occurs for positive $\mu^2$ and negative 
$\lambda$. With these choices, it can be shown that the theory is applicable
only if $|\lambda|\ll 1$. In general, this requires 
fine-tuning: for example, if the dimension-6
operator arises due to exchanges of a heavy weak-singlet scalar, one can
show that $|\lambda|\sim 1$ will also be generated when the scalar is 
integrated out. This should be cancelled by a bare term in the lagrangian. 
However, given the relatively low values of $\Lambda$ interesting for our 
analysis (500~GeV--1~TeV range), the required fine-tuning is numerically
quite modest.\footnote{The appropriate suppression of the quartic coupling
can occur naturally in composite Higgs models~\cite{CH}.}

\begin{figure}[t]
\begin{center}
\includegraphics[width=7cm,keepaspectratio=true]{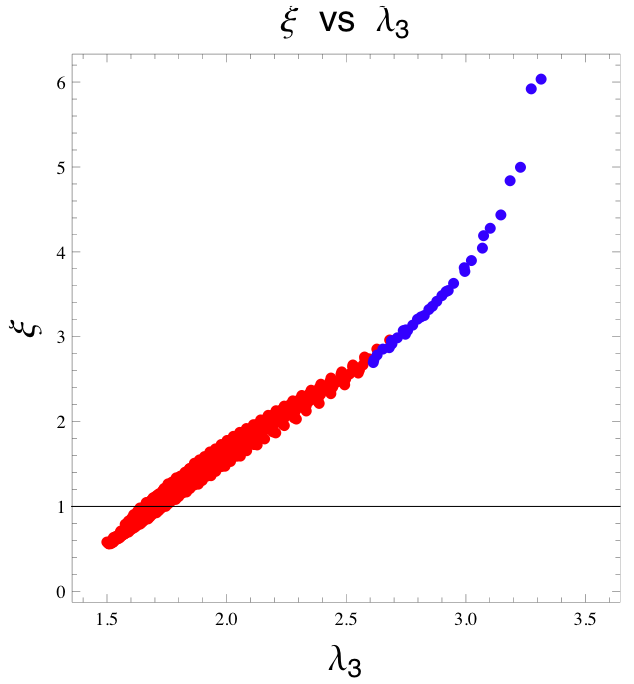}
\hspace{1cm}
\includegraphics[width=7cm,keepaspectratio=true]{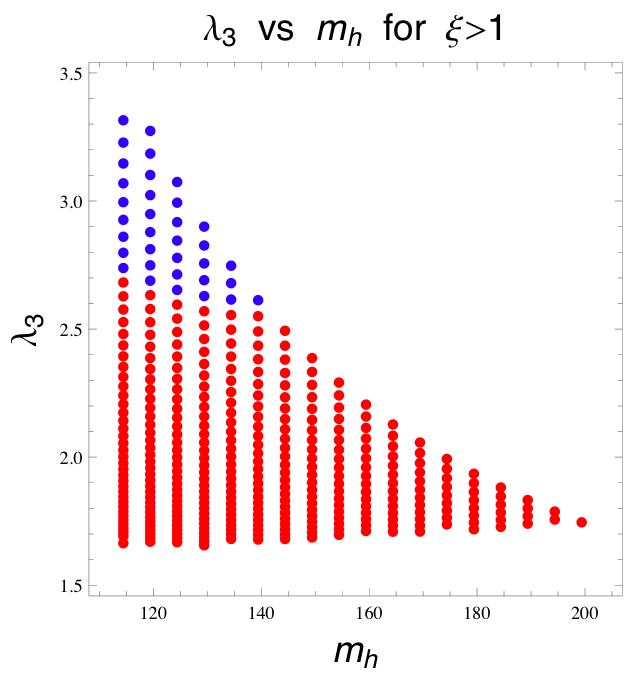}
\caption{SM with an $H^6$ operator. Models with a ``bumpy'' 
zero-temperature Higgs potential are shown in blue, and those without the
bump in red.
(Left panel) The strength of the first-order EWPT $\xi$, defined in 
Eq.~\leqn{xidef}, vs. Higgs cubic self-coupling.  (Right panel)
Higgs cubic self-coupling vs. Higgs mass for points exhibiting a strong
first-order EWPT, $\xi>1$. In both plots, the Higgs self-coupling is 
normalized to the one-loop SM expectation for the same $m_h$.}
\label{fig:h6}
\end{center}
\end{figure} 

We analyzed the dynamics of the EWPT in the GSW model, following the general
framework discussed in Section~\ref{sec:2}. We performed a scan of the
following parameters: $m_h=114.4\ldots 214.4$ GeV, $\Lambda=100\ldots
1000$ GeV. (We did not find any points with strong first-order EWPT for higher 
$\Lambda$ and $m_h$.) Our results for the order of the transition are 
roughly consistent\footnote{Quantitative differences between our analysis and 
those of Refs.~\cite{GSW,DGW} arise due to different approximations made
in each analysis. Our analysis includes the full SM contributions to 
$V^l_0$, $V_T$ and $V_r$, listed in Section~\ref{sec:2}, which were not
included in Ref.~\cite{GSW}. On the other hand, we do not include the
corrections to the one-loop potential and ring terms induced by the 
dimension-6 operator, and do not account for the possibility of delayed 
nucleation. Both effects were included in Ref.~\cite{DGW}. These differences
do not affect the argument of this paper.} with
Refs.~\cite{GSW,DGW}. We also computed 
the Higgs cubic self-coupling (at the one-loop order) for each point in the 
scan. The results are shown in Fig.~\ref{fig:h6}. In the region where there
is a strong first-order EWPT, this model also predicts a large, order-one 
enhancement 
of $\lambda_3$ from its SM value, which will be easily observable at the ILC.
This is easy to understand analytically: at tree level, we have  
\beq
\frac{\lambda_3^{\rm GSW}}{\lambda_3^{\rm SM}}\,=\,1 + \frac{2v^4}{m_h^2
\Lambda^2}\,.
\eeq{tree}
Since strong first-order EWPT can only occur for $\Lambda$ in the 
500 GeV--1 TeV range, the correction term is large for these points. 
(The one-loop corrections to Eq.~\leqn{tree} are at most about 10\% in our 
scan.) In addition, as Fig.~\ref{fig:h6} indicates, there is
a clear and strong correlation between this coupling and the strength of the
EWPT.

\section{A Model with Tree-Level Higgs-Singlet Mixing}
\label{sec:4}

In all examples studied above, the SM-like Higgs field $h$ was the only field
that acquired an expectation value during the EWPT. If other scalar fields
are present, they may change their value in the same transition, so that the
order parameter for the EWPT is effectively a linear combination of $h$ and
other fields. Of particular interest are models with extra gauge-singlet 
scalars, such as the NMSSM, certain Little Higgs models, 
and others. At low temperatures, both singlet and Higgs vevs are generically
non-zero. At high temperatures, the Higgs vev is driven to zero, and although
the singlet vev typically remains non-zero, the EWPT involves changes in
both vevs. Since the zero-temperature singlet potential is not restricted to 
even-degree polynomials by gauge invariance, it may contain cubic terms, 
which can naturally produce large ``bumps'' and
lead to strong first-order EWPT. We will consider a simple model of 
this kind in this section. Our main goal here is to illustrate that the 
connection between the strong first-order EWPT and observable deviations 
of the Higgs self-coupling from the SM value persists in this model.

Many studies of models of this type have appeared in the literature, mostly
in the context of the NMSSM and the nMSSM~\cite{NMSSM}. We will follow the 
conventions of a recent analysis by Profumo, Ramsey-Musolf, and 
Shaughnessy (PRS)~\cite{PRS}, which applies more generally. (See also 
Ref.~\cite{Ahriche}.) Consider a model with a single real scalar $S$ 
added to the SM. The most general renormalizable scalar potential has the form
\beq
V^t(H,S) \,=\, -\mu^2 H^\dagger H + \lambda\left(H^\dagger H\right)^2
+\frac{a_1}{2}H^\dagger H S+\frac{a_2}{2} H^\dagger H S^2+\frac{b_2}{2}S^2+
\frac{b_3}{3}S^3+\frac{b_4}{4}S^4,
\eeq{PRS_Vt_HS}
where $H=(H^+, H^0)$. We assume that the model has a stable vacuum at 
$\left<H^0\right>=v/\sqrt{2}, \left<S\right>=s$. (The conditions for this 
are listed in Ref.~\cite{PRS}.) The physical Higgs fields are obtained by
expanding around this minimum, $H^0=(h+v)/\sqrt{2}$ and $S=s+x$, and
diagonalizing the mass matrix by a rotation
\beqa
h_1&=&\sin\theta~s+\cos\theta~h\,,\CR
h_2&=&\cos\theta~s -\sin\theta~h\,.
\eeqa{h1h2}
The mixing angle is defined such that $|\cos\theta|>1/\sqrt{2}$.  With this 
convention, $h_1$ is more doublet-like and $h_2$ is more singlet-like.  
For positive (negative) $\theta$, the eigenmass $m_2$ is greater than (less 
than) $m_1$. 

We focus on the state $h_1$, which for simplicity we will 
refer to simply as ``the Higgs'', and study the deviation of its cubic 
self-coupling from the SM prediction for a Higgs of the same mass. At tree 
level,
\beqa
\frac{\lambda_3^{\rm PRS}}{\lambda_3^{\rm SM}}&=&
\frac{v}{6 m_1^2}\Bigl( 12 v\lambda\cos^3\theta+3 a_1\cos^2\theta 
\sin\theta+6 a_2(x\cos\theta+v\sin\theta)\cos\theta\sin\theta+ \CR & &
4 b_3\sin^3\theta+12 b_4x\sin^3\theta \Bigr).
\eeqa{PRStree}
Unlike the models studied in section~\ref{sec:2}, no large couplings are 
required to achieve first-order EWPT in this setup. We restrict our 
analysis to the points where all couplings are weak, and
ignore the loop corrections to the relation~\leqn{PRStree}.

\begin{figure}[t]
\begin{center}
\includegraphics[width=7cm,keepaspectratio=true]{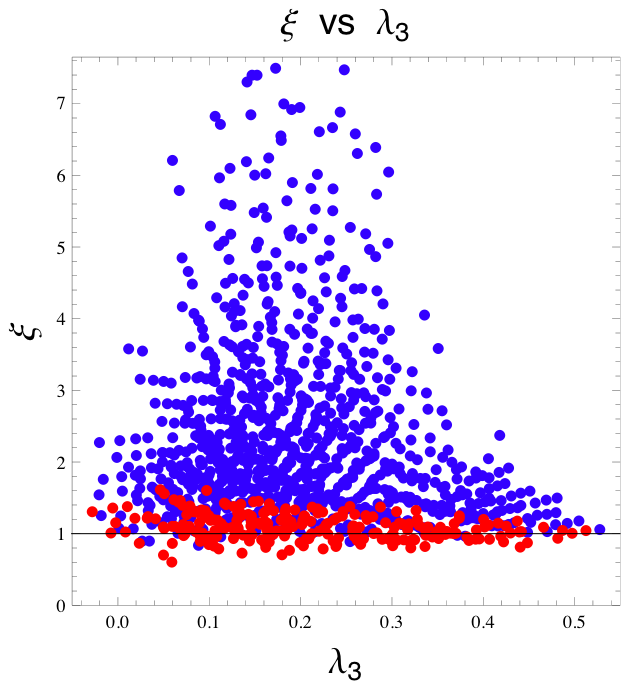}
\hspace{1cm}
\includegraphics[width=7cm,keepaspectratio=true]{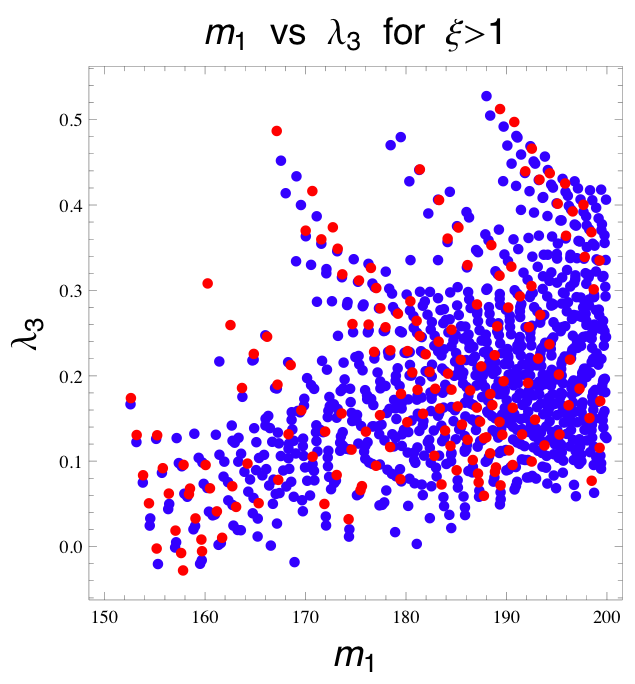}
\caption{SM with a single extra scalar that mixes with the Higgs. Models 
with a ``bumpy'' 
zero-temperature Higgs potential are shown in blue, and those without the
bump in red. (Left panel) The strength of the first-order EWPT $\xi$, 
defined in 
Eq.~\leqn{xidef}, vs. Higgs cubic self-coupling. (Right panel)
Higgs cubic self-coupling vs. Higgs mass for points exhibiting a strong
first-order EWPT, $\xi>1$. In both plots, the Higgs self-coupling is 
normalized to the tree-level SM expectation for the same $m_h$.}
\label{fig:svevblob}
\end{center}
\end{figure} 

It is straighforward to compute the finite-temperature one-loop potential in
this model, and analyze the dynamics of the EWPT numerically. We used the 
high-temperature approximation for $V^t$, expanding it up to linear order in 
$T$. This approximation greatly speeds up the numerical analysis. We 
performed 
a partial scan of the parameter space of the model, setting $a_2=0$ and 
scanning $a_1=-150\ldots-50$ GeV, $b_3=-150\ldots-50$ GeV, 
$\lambda=0.15\ldots
0.25$, $x=150\ldots250$ GeV, and $b_4=0.5$. This parameter region includes 
a large set of points with strong first-order EWPT, and lies roughly 
within the range consistent with precision electroweak 
constraints\footnote{Note, however, that other new weak-scale states can 
contribute to precision electroweak observables and modify the fits, without
changing the dynamics of the EWPT. A SM-like Higgs in the
$300-800$ GeV mass range is allowed if new physics produces a positive 
contribution to the T parameter. This is the case, for example, in the 
LHT model~\cite{HMNP}.} determined in~\cite{PRS}. The results of this scan 
are illustrated in Fig.~\ref{fig:svevblob}. The models with strong 
first-order 
EWPT predict a suppression of the Higgs cubic self-coupling relative to 
the SM value, by a factor of 2 or more. The sign of the effect, however, 
is not unique in this model: for example, we found points with $a_2\not=0$
which predict a strong first-order EWPT and an enhanced Higgs 
self-coupling. The important point is that large, order-one deviations in 
$\lambda_3$ from the SM are typical. This is not surprising, 
since both these deviations and the possibility of strong first-order
EWPT are due to same new tree-level interactions.

\begin{figure}[t]
\begin{center}
\includegraphics[width=7cm,keepaspectratio=true]{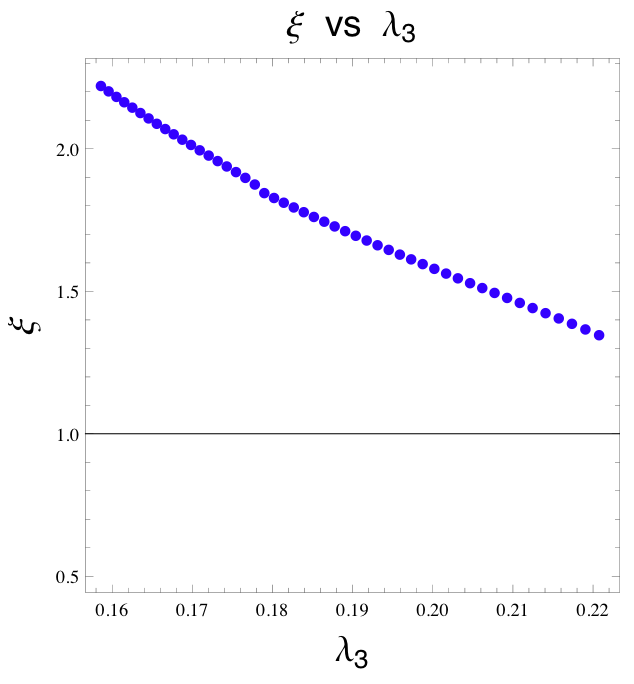}
\caption{The strength of the first-order EWPT $\xi$ vs. Higgs cubic 
self-coupling, in the model of Eq.~\leqn{PRS_Vt_HS}. Only the 
mixing parameter, $a_1$, is varied. The Higgs self-coupling is 
normalized to the one-loop SM expectation for the same $m_h$.}
\label{fig:sveva1}
\end{center}
\end{figure}

The corellation between the size of this effect and the strength of the 
EWPT is less clear than in the examples of sections~\ref{sec:2} 
and~\ref{sec:3}. This is simply due to a large number of parameters that 
affect $\lambda_3$, but have only a marginal effect on $\xi$. If those 
parameters are fixed, and, for example, only the Higgs-singlet mixing 
parameter $a_1$ is varied, a clear corellation between $\lambda_3$ and $\xi$
emerges, as seen in Fig.~\ref{fig:sveva1}. (In this figure we fixed 
$b_3=-80$ GeV, $\lambda=0.20$, $x=150$ GeV, $b_4=0.5$, $a_2=0$, and scanned 
$a_1=-130\ldots-105$ GeV.) 

Unlike the models of sections~\ref{sec:2} and~\ref{sec:3}, where the 
non-standard Higgs cubic coupling might provide the first experimental hint
of the existence of extra singlet scalars at the weak scale, the
model studied here will provide a variety of deviations in the
Higgs production cross section and branching ratios from the SM due to the 
singlet admixture~\cite{PRS}. Many such deviations will probably be observed 
before the Higgs self-coupling measurement is available, especially if 
$\lambda_3$ is suppressed. Nevertheless, it is interesting that the general 
pattern of large deviations in $\lambda_3$ in all models exhibiting a
strong first-order EWPT persists in this class of models.

\if  
we fixed $b_3=-80$ GeV, $\lambda=0.20$, $x=150$ GeV, $b_4=0.5$, $a_2=0$, and scanned $a_1=-130\ldots-106$ GeV.  The results are presented in Fig.~\ref{fig:sveva1}.  This scenario demonstrates a direct proportionality between $a_1$ and $\xi$, but an inverse proportionality between $a_1$ and $\lambda_3$.
 
The origin of this supression is as follows. The terms in the potential 
involving only the doublet components of the $h_1$ produce a positive 
contribution to $\lambda_3$, as in the SM. (This contribution reduces to the 
SM value in the no-mixing limit.) The term involving doublet-singlet mixing, 
as well as the $S^3$ term, produce a negative contribution for $\theta>0$, 
since their coefficients $a_2$ and $b_3$ are negative. Note that the sign of
$a_2$ and $b_3$ is effectively determined by the requirement of strong
first-order EWPT: it is these terms that introduce the desired ``bump'' in the 
potential. 
\fi

\section{Conclusions}

In this paper, we investigated the link between the dynamics of the EWPT 
and the deviation of the cubic Higgs boson self-coupling 
in extensions of the SM. We studied this link in a series of toy models. 
Our toy models implement three mechanisms by which the weak-scale new physics
could turn the second-order adiabatic transition predicted in the pure SM
into a first-order transition. In the first set of models, the dynamics is 
determined by large quantum corrections to the Higgs potential, both zero- 
and finite-temperature, from loops of new particles (possibly, but not 
necessarily, gauge singlets) with strong coupling to the Higgs. In the 
second model, the first-order transition is due to the presence of a 
nonrenormalizable operator in the Higgs potential. In the third model, the
crucial new feature is the tree-level mixing between the Higgs and an
extra scalar singlet. In all cases, we found that models that exhibit a
strong first-order EWPT predict large deviations of the Higgs   
cubic coupling from the Standard Model value. Order-one deviations are
typical. The models of the first and second classes typically predict an 
enhancement in the Higgs cubic coupling, while in the models of the third
class both signs are equally likely. The only cases where the cubic coupling
does {\it not} deviate significantly from the SM prediction are
due to accidental cancellations among unrelated contributions to this
coupling. Thus, we conclude that a measurement of the Higgs cubic coupling consistent with the SM prediction would strongly disfavor (although not completely rule out) a first-order electroweak phase transition. A discovery of a substantial deviation of this coupling from the SM would keep the possibility of a first-order transition open, but would not conclusively prove it, as discussed at the end of section~\ref{singleS}.

While a measurement of the Higgs cubic coupling is very difficult at the
LHC, a 20\%-level measurement appears to be within the reach of the proposed
luminosity or energy upgrades of the LHC, as well as the ILC. 
Based on the analyis, we conclude that such a measurement could serve as
a sensitive probe of the dynamics of the electroweak phase tranition. If
no deviation from the SM is observed, it would be very unlikely that the
transition is first-order. Note that at least some of the sources of the 
first-order transition are very difficult to exclude by any other means: 
for example, a gauge-singlet scalar that does not mix with the Higgs 
at tree level but has a strong quadratic coupling, as in the model of 
section~\ref{singleS}, is essentially impossible to detect directly as 
long as the decay $h\to SS$ is kinematically forbidden. Therefore, even with
a large body of knowledge about the BSM particles and their couplings 
from the LHC and ILC experiments, the question of the order of the phase
transition would be very difficult to settle definitively. A measurement
of the Higgs cubic coupling would be crucial in addressing this issue.

It is exciting that the next generation of collider experiments will 
likely contribute to our understanding of cosmology. One example 
is the possibility that the dark matter particle will be discovered, and
its propserties measured, at the LHC and the ILC. This connection has 
attracted much attention in the literature, and has been analyzed both in
specific models of new physics~\cite{Peskin}, and in a model-independent 
framework~\cite{MIDM}. Understanding the dynamics of the electroweak 
phase transition is another open question in the early universe cosmology
where the LHC and ILC experiments will have an important role to play. 
In this paper, we studied one simple experimentally observable quantity 
that is tightly correlated with the order of the transition. It is also 
possible that a combination of mass and coupling measurements could be 
used to completely reconstruct the Higgs potential, providing a much more
detailed picture of the transition. It would be interesting to understand
what quantities would need to be measured, and how precise the measurements
would need to be. 

\vskip1cm

\noindent{\large \bf Acknowledgments} 

We are grateful to Sohang Gandhi, Christophe Grojean and Geraldine Servant for 
useful discussions. We are also very grateful to David Rainwater for a 
communication regarding the prospects for a measurement of $\lambda_3$ at 
the LHC. This research is supported by the NSF grant PHY-0355005.


\begin{thebibliography}{99}

\bibitem{EWPT}
  D.~A.~Kirzhnits,
  JETP Lett.\  {\bf 15}, 529 (1972)
  [Pisma Zh.\ Eksp.\ Teor.\ Fiz.\  {\bf 15}, 745 (1972)];
D.~A.~Kirzhnits and A.~D.~Linde,
  Phys.\ Lett.\  B {\bf 42} (1972) 471.

\bibitem{EWbaryo}
  M.~E.~Shaposhnikov,
  Nucl.\ Phys.\  B {\bf 287}, 757 (1987);
  Nucl.\ Phys.\  B {\bf 299}, 797 (1988).
For reviews of more recent developments, see
A.~Riotto and M.~Trodden,
  Ann.\ Rev.\ Nucl.\ Part.\ Sci.\  {\bf 49}, 35 (1999)
  [arXiv:hep-ph/9901362];
  M.~Dine and A.~Kusenko,
  Rev.\ Mod.\ Phys.\  {\bf 76}, 1 (2004)
  [arXiv:hep-ph/0303065].

\bibitem{Grojean:2006bp}
  C.~Grojean and G.~Servant,
  Phys.\ Rev.\  D {\bf 75}, 043507 (2007)
  [arXiv:hep-ph/0607107];
  C.~Caprini, R.~Durrer and G.~Servant,
  arXiv:0711.2593 [astro-ph].

\bibitem{GSW}
  C.~Grojean, G.~Servant and J.~D.~Wells,
  Phys.\ Rev.\  D {\bf 71}, 036001 (2005)
  [arXiv:hep-ph/0407019].

\bibitem{2HDM}
  S.~Kanemura, Y.~Okada and E.~Senaha,
  Phys.\ Lett.\  B {\bf 606}, 361 (2005)
  [arXiv:hep-ph/0411354]; 
``Electroweak baryogenesis and the triple Higgs boson coupling,''
{\it In the Proceedings of 2005 International Linear Collider Workshop (LCWS 2005), Stanford, California, 18-22 Mar 2005, pp 0704}
  [arXiv:hep-ph/0507259];
   S.~W.~Ham and S.~K.~Oh,
  [arXiv:hep-ph/0502116].

\bibitem{MSSM}
For reviews, see 
  H.~E.~Haber and G.~L.~Kane,
  Phys.\ Rept.\  {\bf 117}, 75 (1985);
  S.~P.~Martin,
  arXiv:hep-ph/9709356.

\bibitem{LH}
For reviews, see 
 M.~Schmaltz and D.~Tucker-Smith,
  Ann.\ Rev.\ Nucl.\ Part.\ Sci.\  {\bf 55}, 229 (2005)
  [arXiv:hep-ph/0502182];
  M.~Perelstein,
  Prog.\ Part.\ Nucl.\ Phys.\  {\bf 58}, 247 (2007)
  [arXiv:hep-ph/0512128].

\bibitem{AH}
  G.~W.~Anderson and L.~J.~Hall,
  Phys.\ Rev.\  D {\bf 45}, 2685 (1992).

\bibitem{WDJ}
  L.~Dolan and R.~Jackiw,
  Phys.\ Rev.\  D {\bf 9}, 3320 (1974);
 S.~Weinberg,
  Phys.\ Rev.\  D {\bf 9}, 3357 (1974).

\bibitem{Carrington:1991hz}
  M.~E.~Carrington,
  Phys.\ Rev.\  D {\bf 45}, 2933 (1992).
  
\bibitem{WW}
  E.~J.~Weinberg and A.~q.~Wu,
  Phys.\ Rev.\  D {\bf 36}, 2474 (1987).

\bibitem{BPR1}
  U.~Baur, T.~Plehn and D.~L.~Rainwater,
  Phys.\ Rev.\ Lett.\  {\bf 89}, 151801 (2002)
  [arXiv:hep-ph/0206024].

\bibitem{BPR2}
  U.~Baur, T.~Plehn and D.~L.~Rainwater,
  Phys.\ Rev.\  D {\bf 67}, 033003 (2003)
  [arXiv:hep-ph/0211224].

\bibitem{BPR3}
  U.~Baur, T.~Plehn and D.~L.~Rainwater,
  Phys.\ Rev.\  D {\bf 69}, 053004 (2004)
  [arXiv:hep-ph/0310056].

\bibitem{Djouadi:2007ik}
  A.~Djouadi, J.~Lykken, K.~Monig, Y.~Okada, M.~J.~Oreglia and S.~Yamashita,
  arXiv:0709.1893 [hep-ph].
  
\bibitem{DT}
  S.~R.~Coleman and E.~Weinberg,
  Phys.\ Rev.\  D {\bf 7}, 1888 (1973).

\bibitem{GGS}
  W.~D.~Goldberger, B.~Grinstein and W.~Skiba,
  arXiv:0708.1463 [hep-ph].

\bibitem{EQ}
  J.~R.~Espinosa and M.~Quiros,
  Phys.\ Rev.\  D {\bf 76}, 076004 (2007)
  [arXiv:hep-ph/0701145].

\bibitem{HK}
  S.~J.~Huber and T.~Konstandin,
  arXiv:0709.2091 [hep-ph].

\bibitem{DGW}
  C.~Delaunay, C.~Grojean and J.~D.~Wells,
  arXiv:0711.2511 [hep-ph].

\bibitem{CH}
  G.~F.~Giudice, C.~Grojean, A.~Pomarol and R.~Rattazzi,
  JHEP {\bf 0706}, 045 (2007)
  [arXiv:hep-ph/0703164].

\bibitem{NMSSM}
See, for example: 
  M.~Pietroni,
  Nucl.\ Phys.\  B {\bf 402}, 27 (1993)
  [arXiv:hep-ph/9207227];
S.~J.~Huber and M.~G.~Schmidt,
  Nucl.\ Phys.\  B {\bf 606}, 183 (2001)
  [arXiv:hep-ph/0003122];
A.~Menon, D.~E.~Morrissey and C.~E.~M.~Wagner,
  Phys.\ Rev.\  D {\bf 70}, 035005 (2004)
  [arXiv:hep-ph/0404184];
  C.~Balazs, M.~S.~Carena, A.~Freitas and C.~E.~M.~Wagner,
  JHEP {\bf 0706}, 066 (2007)
  [arXiv:0705.0431 [hep-ph]].


\bibitem{PRS}
  S.~Profumo, M.~J.~Ramsey-Musolf and G.~Shaughnessy,
  JHEP {\bf 0708}, 010 (2007)
  [arXiv:0705.2425 [hep-ph]].

\bibitem{Ahriche}
  A.~Ahriche,
  Phys.\ Rev.\  D {\bf 75}, 083522 (2007)
  [arXiv:hep-ph/0701192].

\bibitem{HMNP}
  J.~Hubisz, P.~Meade, A.~Noble and M.~Perelstein,
  JHEP {\bf 0601}, 135 (2006)
  [arXiv:hep-ph/0506042].

\bibitem{Peskin}
See, for example, 
 E.~A.~Baltz, M.~Battaglia, M.~E.~Peskin and T.~Wizansky,
  Phys.\ Rev.\  D {\bf 74}, 103521 (2006)
  [arXiv:hep-ph/0602187].

\bibitem{MIDM}
  A.~Birkedal, K.~Matchev and M.~Perelstein,
  Phys.\ Rev.\  D {\bf 70}, 077701 (2004)
  [arXiv:hep-ph/0403004];
  C.~Bartels and J.~List,
  arXiv:0709.2629 [hep-ex].

\end{thebibliography}
\end{document}